\documentclass[twocolumn,showpacs,preprintnumbers,amsmath,amssymb,prl]{revtex4}
\usepackage{graphicx}
\usepackage{dcolumn}
\usepackage{bm}
\usepackage{amssymb}
\usepackage{amsmath}
\usepackage{epsf}
\usepackage{amssymb}
\usepackage{amsmath}
\usepackage{epsf}
\usepackage{graphics}
\usepackage{verbatim}
\def\be{\begin{equation}}
\def\ee{\end{equation}}

\def \rm {\mathrm}

\begin{document}

\title{Spin waves in diluted magnetic quantum wells}

\author{P. M. Shmakov$^{2}$}
\author{A. P. Dmitriev$^{1,2}$}
\author{V. Yu. Kachorovskii$^{1,2}$}
\affiliation{
$^{1}$Institut f\"ur Nanotechnologie, Forschungszentrum Karlsruhe,
76021 Karlsruhe, Germany
\\
$^{2}$A.F.Ioffe Physico-Technical Institute,
194021 St.~Petersburg, Russia
}

\date{\today}

\begin{abstract} {We study collective spin excitations in two-dimensional diluted magnetic semiconductors, placed into external magnetic field.  Two coupled modes of the spin waves (the electron and ion modes) are found to exist in the system along with a number of the ion spin excitations decoupled from the electron system. We calculate analytically the spectrum of the waves taking into account the exchange interaction of itinerant electrons both with each other and with electrons localized on the magnetic ions. The interplay of these interactions leads to a number of intriguing phenomena including tunable anticrossing of the modes and a  field-induced change in a sign of the group velocity of the ion mode.} \end{abstract} \maketitle

  Diluted magnetic semiconductors (DMS)  have recently been the  subject of great  interest \cite{1,2} due to their  potential in combining magnetic and semiconductor properties in a single material. The DMS are formed  by  replacing  of  cations in ordinary semiconductors  with magnetic ions, typically Mn ions. Strong exchange interaction between the  itinerant electrons and the  electrons localized  on d-shells of the magnetic ions leads to a number of  remarkable features of the DMS. In particular, it results in the effective indirect  interaction between  the ion spins thus promising for creating room-temperature ferromagnetic systems that may offer  advantages of semiconductors. It also dramatically  enhances the effective coupling of the itinerant electrons with the external magnetic field. In contrast to conventional GaAs/GaAlAs systems, where small values of $g-$factor prevents manipulation  of the spin degree of freedom, the  giant electron Zeeman splitting  arising in the DMS as a manifestation of the exchange interaction can be on the order of the Fermi energy \cite{3,4}, offering a wide range of  spintronics applications.

\begin{figure}[ht!]
 \leavevmode \epsfxsize=3.65cm \centering{\epsfbox{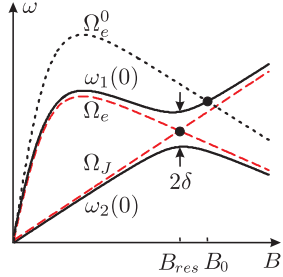}}
\caption{Anticrossing of the electron and ion spin precession frequencies $\Omega_e$ and $\Omega_J$ at "resonant" magnetic field $B = B_{res}$;
$\Omega_e^0$ is the edge of the Stoner continuum.}
\label{fig1}
\end{figure}

Here we discuss spin excitations in the  two-dimensional DMS. Our studies are motivated by recent experiments \cite{5,6,7} and a theoretical discussion \cite{7,10,8} focused on the  spin dynamics in  diluted magnetic Cd$_{1-x}$Mn$_x$Te quantum wells placed into the magnetic field \cite{mm}. In Ref. \onlinecite{5},  the spectrum of the spin waves, $\omega(k),$ was measured. Only {\it one}  excitation  mode was observed.  It was demonstrated that the excitations exist in a finite range of wavelengths, $k<k_{m},$ and their group velocity   is negative: $d\omega(k)/dk <0$. The experimental data were interpreted \cite{5,10} in terms of conventional spin waves in the Fermi liquids \cite{11}, while $ k_{m}$  was attributed to the edge of  the Stoner continuum of the single-particle spin excitations. Such interpretation implies that the only effect of the magnetic ions on the electron spin waves is the strong renormalization of the electron Zeeman splitting. However, more recent experimental observations \cite{6,7} supported by   theoretical studies \cite{7,8}   appear to be   in disagreement with this conclusion. Indeed, in Refs.~\cite{6,7} {\it two} modes of the collective  homogeneous ($k=0$) spin excitations were observed in Cd$_{1-x}$Mn$_x$Te wells. The modes were identified \cite{6,7,8} as   the  spin excitations of delocalized electrons (the electron mode) and the  electrons  on d-shells of Mn  ions (the ion mode). The dependencies of the  frequencies $\omega_{1,2}(0)$  of observed modes on $B$ are shown schematically  in Fig.~\eqref{fig1}. The most important observation  is the anticrossing of the modes which occurs at a certain  "resonant" value of magnetic field $B=B_{res}$ \cite{6,7}. As it was also shown in Ref.~\cite{7},  other types of the   homogeneous spin modes  may exist in the system corresponding to excitations of  the ion spins decoupled from the spins of the itinerant electrons.

In this paper, we develop a theory of the  spin waves in diluted magnetic quantum  wells placed into magnetic field.  We study analytically  two  collective  modes    which correspond to coupled propagation of the electron and ion spin excitations. We also discuss the ion modes  decoupled from the electron system. To describe the homogeneous spin oscillations  ($k=0$), it is sufficient to take into account only one type of exchange interaction: the  interaction of  the itinerant electrons with electrons localized on the Mn ions.  The thus obtained results     coincide with those presented  in Refs.~\cite{6,7}. For $k\neq 0,$ the   exchange interaction  between delocalized electrons  comes into play. Our main purpose is to demonstrate that the simultaneous presence of two types of exchange interaction gives rise to interesting phenomena  the most remarkable one is  the magnetic-field-driven  anticrossing of  the spin modes. In contrast to the case $k=0,$   the anticrossing can take place in a wide range of the magnetic fields and may be tuned by the field to occur at a certain value of $k$  (see Fig.~2).

We consider the 2D degenerate electron gas  interacting with  randomly placed magnetic ions.  The electrons are located in the $\mathbf r=(x,y)$ plane and  occupy the lowest level in the well. The ions are distributed  homogeneously with the 2D concentration $n_J$, which is  assumed to be much higher than the electron concentration $n_e.$ The magnetic field  is  applied parallel to the well plane ($\mathbf B \parallel \mathbf e_x$). The field leads to the Zeeman splitting of the electron and ion spin levels with energies $\hbar \omega_e$ and $\hbar \Omega_J,$ respectively, while the orbital motion remains intact. The Hamiltonian of electron-ion exchange interaction reads $\hat H_{Je} = -\alpha/2\sum_{k}\boldsymbol{\hat \sigma}\mathbf{\hat J}_k \delta(\mathbf r - \mathbf r_k)|\Psi(z_k)|^2,$ where $\boldsymbol{\hat \sigma}$ is the Pauli matrix vector,  $\mathbf{\hat J}_k$ are the  spin operators of  the ions located at the points $\mathbf R_k=(\mathbf r_k, z_k),$ and   $\Psi(z) = \sqrt{2/a}\sin(\pi z/a)$ is a wave function of the lowest level in a rectangular well of width $a.$ Since $n_J \gg n_e$, the distance between the ions is much smaller than the electron wave length and the mean-field approximation is applicable. In this approximation, we first replace   $\boldsymbol{\hat \sigma}\mathbf{\hat J}_k$  with $\langle \boldsymbol{\hat \sigma}\rangle \mathbf{\hat J}_k +\boldsymbol{\hat \sigma}\langle \mathbf{\hat J}_k \rangle,$ where  averaging is taken over density matrix of the system $\hat\rho$. After such  decoupling one may  search the solution of the quantum Liouville   equation for  $\hat \rho$ as a product of the electron  and ion density matrices: 
$\hat \rho = \hat \rho_e(\mathbf r,\mathbf r', t)\prod_k \hat \rho_k(t)$.
The averaged spins of the electrons and ions are given by $\mathbf{s}_0(\mathbf r,t) = n_e^{-1} \int \mathbf s(\mathbf r, \mathbf p,t) d^2 \mathbf p/(2\pi\hbar)^2 $, $ \mathbf{ J}_k(t) = \rm{Tr}(\hat \rho_k(t) \mathbf{\hat J}_k )   ,$ where $\mathbf s(\mathbf r, \mathbf p,t)=  \rm{Tr}(  \boldsymbol {\hat \sigma} \hat f)/2 $ is the electron spin density and $ \hat f =\hat f(\mathbf r,\mathbf p, t)$ is the Wigner function corresponding to $\hat \rho_e.$ As a next step, we replace $ \mathbf{ J}_k(t) $ with a smooth function $\mathbf J(\mathbf r,z,t).$ Doing so, one finds   the electron spin precession frequency in the ion-induced exchange field
\begin{equation}
 \boldsymbol\omega_{eJ} (\mathbf r,t)  = {\alpha n_J \bar {\mathbf J}(\mathbf r,z,t)}/{\hbar a} ,
  \label{omegaeJ}
  \end{equation}
[here $\bar{ \mathbf J} (\mathbf r,t)= \int dz |\Psi(z)|^2\mathbf{ J}(\mathbf r,z,t)$], and  the local frequency of the ion spin precession
\begin{equation}
  \boldsymbol\omega_{Je}(\mathbf r,z,t) = {\alpha n_e \mathbf{s}_0(\mathbf r,t)|\Psi(z)|^2} /{\hbar}.
   \label{omegaJe}
  \end{equation}
In addition to the electron-ion exchange interaction, we take into account the isotropic ferromagnetic electron-electron exchange interaction by adding the following term \cite{11}
\begin{equation}
 \boldsymbol \omega_{ee}(\mathbf r,t) = -{2G}n_e \nu^{-1} \mathbf{s}_0(\mathbf r,t),
  \label{omega_ee}
  \end{equation}
  to the  electron spin precession  frequency. Here  $G<0$ is the  interaction constant (we assume $|G| <1$), $\nu=m/2\pi\hbar^2$, and $m$ is the electron effective mass.

The  equilibrium electron spin density,
   $\mathbf s^{\text{eq}}(\mathbf p) = \left[n_\uparrow({\epsilon })- n_\downarrow({\epsilon )}\right]{\mathbf e_x}/2$
 (here \mbox{$n_{\uparrow \downarrow}(\epsilon) = \left\{\exp[(\epsilon \mp  {\hbar \Omega_e^0}/{2} - E_F)/T]+1 \right\}^{-1}$ and }  $\epsilon=p^2/2m$), is expressed  via the effective Zeeman splitting $ \hbar\Omega_e^0$.
    The  frequency $\Omega_e^0=\omega_e+\omega_{eJ}+\omega_{ee}$ is found self-consistently from  Eqs.~\eqref{omegaeJ} and  \eqref{omega_ee}:
  \begin{equation}\Omega_e^0 = {\Omega_e}/{(1+G)} , \end{equation}
where
\begin{equation}\Omega_e = \omega_e +\alpha n_J  J_x ^{\rm {eq}}/a \hbar, \end{equation}
is the effective electron Zeeman splitting renormalized by exchange interaction with the ions.
In deriving these equations,
 $\mathbf J(\mathbf r,z)$ was substituted with the equilibrium ion polarization,
   $\mathbf J^{\text{eq}} =   J_x^{\text{eq}}  \mathbf e_x=  \frac52 B_{5/2}(\hbar\Omega_J/T\mathbf) \mathbf e_x$, where $B_J(x)$
   is the Brillouin function. We also assumed
    that the equilibrium exchange field acting on the ions is small,
   $\alpha n_e  s_x^{\text{eq}}/a \approx \alpha \hbar \Omega_e^0n_e/4E_F a \ll \hbar \Omega_J,$
         which implies that the equilibrium ion polarization is  not affected by exchange interaction.
   In contrast, the electron Zeeman splitting  is strongly enhanced  due to high ion concentration, so that  $\Omega_e \gg |\omega_e|$
   ($\omega_e<0,$ because of the  negative electron $g$-factor, which explains non-monotonic dependence of $\Omega_e$ on $B$ shown in Fig.~1  \cite{6,7}).

 The  out-of-equilibrium spin dynamics
   can be described by   the Landau-Silin equation \cite{12} for the electrons
    and   the Bloch equation for the ions:
   \begin{align}
&\frac{\partial \hat f}{\partial t} + \frac{\mathbf p}{m}\nabla \hat f - \frac12 \left\{\frac{\partial \hat f}{\partial \mathbf p},\frac{\partial \hat \varepsilon}{\partial \mathbf r}\right\} +\frac i\hbar [\hat \varepsilon, \hat f] = 0,
\label{dfdt}
\\
&\frac{\partial \mathbf J}{\partial t} + [\Omega_J \mathbf e_x + \boldsymbol \omega_{Je}] \times \mathbf J = 0.
\label{dJdt}
\end{align}
Here $[\cdots]$ and $\{\cdots \}$ stand for
the commutator and the anticommutator, respectively, and $\hat\varepsilon = - \hbar [\omega_e \mathbf e_x +
\boldsymbol\omega_{eJ} + \boldsymbol\omega_{ee}] \boldsymbol{\hat \sigma}/2.$
For $\hbar \Omega_e^0 \ll E_F,$ Eqs.~\eqref{dfdt} and \eqref{dJdt} give the following system of  equations for  the perpendicular (with respect to $\mathbf B$) components of the electron and ion spins
 \begin{align}
&\frac{\partial s}{\partial t} + (v_F \mathbf n \nabla + i\Omega_e^0)(s+Gs_0) = \delta_1(i\bar{J} + \xi \mathbf n\nabla \bar{J}),
\nonumber
\\
&\frac{\partial \bar{J}}{\partial t} + i\Omega_J \bar{J} = i\delta_2 s_0.
\label{J}
\end{align}
 Here $s = s_y+is_z, \bar{J} = \bar{J}_{y}+i\bar{J}_{z},$
      $v_F$ is Fermi velocity, $\mathbf n = (\cos \varphi, \sin \varphi),$  $\varphi$ is the velocity angle in the well plane,
       $\delta_1 = \alpha n_J  s_x^{\text{eq}}  /\hbar a$,
 $\delta_2 =  3\alpha n_e   J_x ^{\text{eq}} /2\hbar a$,  $  \xi={v_F}/{\Omega_e^0}$
 and $s_0 = \int_0^{2\pi} s d\varphi/2\pi$ \cite{13}.
      After Fourier transform of Eqs.~\eqref{J},  one can find
             the dispersion equation for the collective modes
    $$    \sqrt{1- \frac{v_F^2k^2}{(\omega-\Omega_e^0)^2}}
                  =\frac{\omega}{\omega-\Omega_e^0}\frac{\delta^2 + G\Omega_e^0(\omega-\Omega_J)}{\delta^2 + \Omega_e(\omega-\Omega_J)},$$
         where $\delta = \sqrt{\delta_1\delta_2}.$
     For $k=0,$  we get  \cite{7,8}, $\omega_{1,2}(0)= (\Omega_e+\Omega_J)/2  \pm\sqrt {(\Omega_e-\Omega_J)^2/4+\delta^2}.$  The anticrossing occurs
          when $\Omega_e(B) =\Omega_J(B).$  We see that the constant $G$ drops out from $\omega_{1,2}(0)$.
\begin{figure}[ht!]
\leavevmode  \epsfxsize=3.65cm\centering{\epsfbox{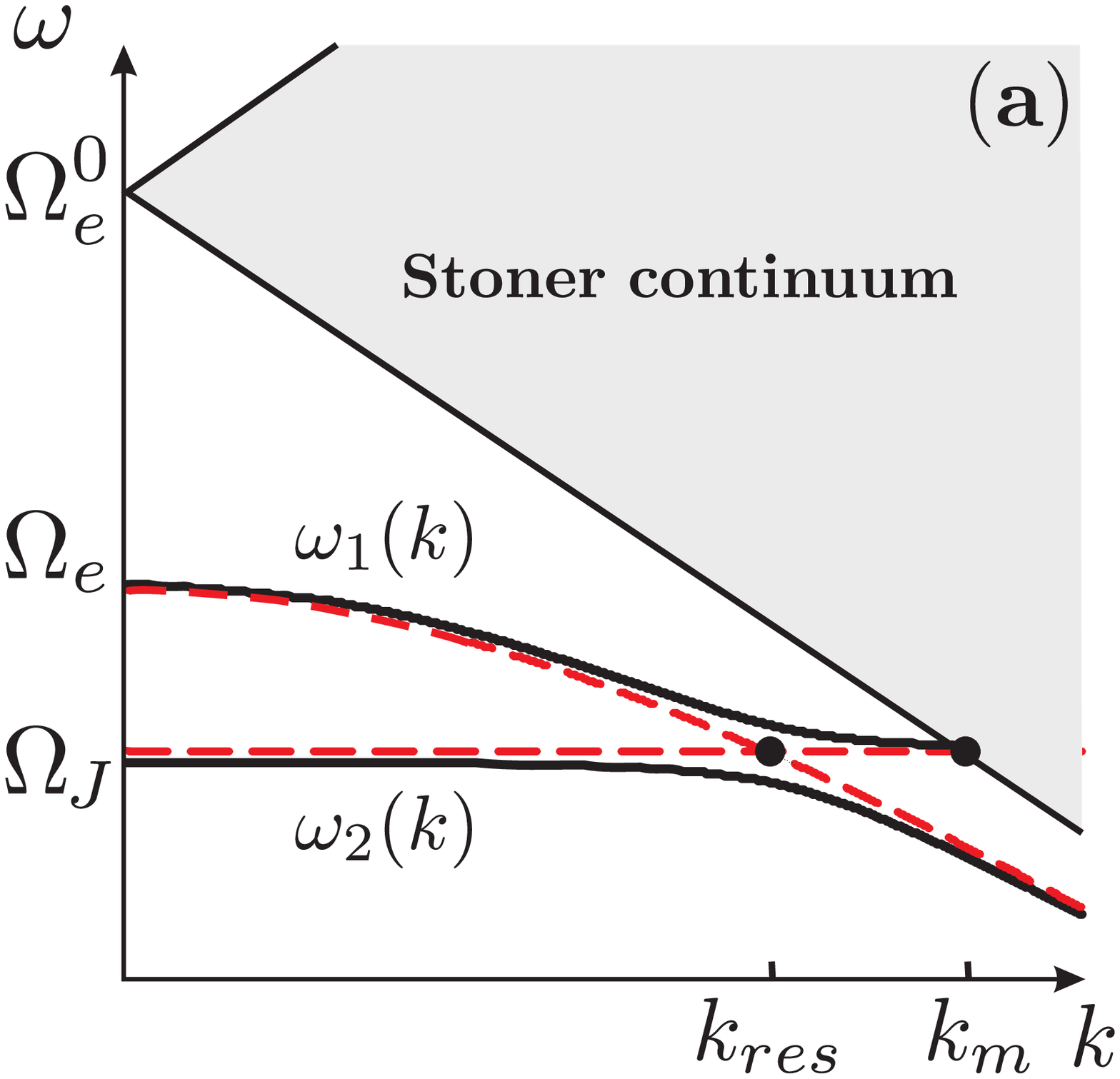}}
\leavevmode \epsfxsize=3.65cm\centering{\epsfbox{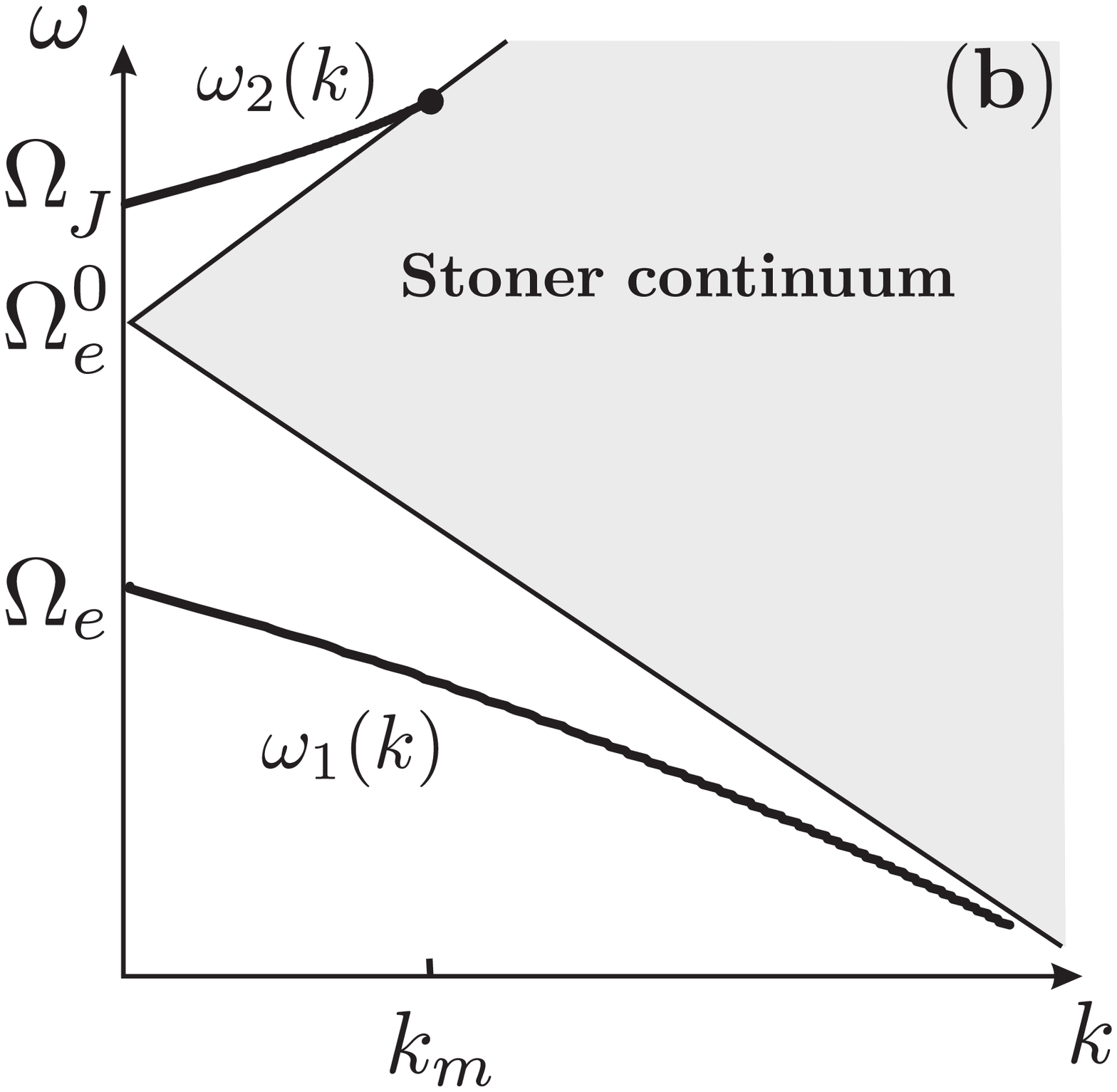}}
\caption{(a) Anticrossing of the two collective spin modes for $B<B_{res}$;
(b) Sign inversion of the group velocity of the ion mode for $B>\tilde B$.}
\end{figure}
    In contrast, the  dispersion of the collective modes strongly depends on the relation between $|G|$ and the dimensionless parameter $\delta/\Omega_e.$     For $\delta/\Omega_e \ll |G|$  (this  was the case in the experiment \cite{5}), the anti-crossing   occurs  for $B<B_{res},$ when $\Omega_e > \Omega_J$ (see Fig.~2a). To see this,  one may  consider  the case $\delta = 0$ (coupling between transverse components of electron and ion spins  is turned off) as a first approximation. This   approximation   was implicitly used in Ref.~\cite{10}. For $\delta=0,$  there are two  branches of the spectrum (dashed lines in Fig.~2a), corresponding respectively to the Fermi-liquid spin waves with negative dispersion and the dispersionless excitations of the ion spins.      Importantly,  for $B < B_{res}$ these two branches intersect each other. Turning on a finite coupling between modes, $\delta \neq 0,$ results in the anticrossing, which, for $B$ close to $B_{res},$ occurs at the point \begin{equation} k_{res} = {\sqrt{2|G|\Omega_e(\Omega_e-\Omega_J)}}/{v_F(1+G)}. \label{anti} \end{equation}
      Remarkably, $k_{res}$ depends on $B$, so that the anticrossing position may be tuned by the external field. The splitting between the modes for $k\approx k_{res}$ is given by $\omega_1(k_{res}) - \omega_2(k_{res})\approx 2\delta.$

    As seen from Fig.~2a the upper branch of the spectrum at a certain wave vector $k_m$ reaches the Stoner continuum (single-particle excitations), which is defined by inequality $|\omega - \Omega_e^0|\leq v_Fk$.  For $k>k_m$ the corresponding ion-type excitations slowly decay in time due to weak  exchange coupling  with  the system of itinerant electrons.  This decay is similar to the well-known Landau damping in plasma \cite{12}, so that the decay rate $\gamma$  is calculated in a quite analogous way, yielding
         \begin{equation}
  \gamma\approx \frac{\delta^2\Omega_J v_F\sqrt{k^2-k_m^2}}{\Omega_e^0[(1+G)^2v_F^2(k^2-k_m^2)+G^2\Omega_J^2]}.
 \end{equation}
As a function of $k$, $\gamma$ has a maximum. The maximal value is given by $\gamma_{\rm{max}}=\delta^2/2|G|\Omega_e \approx 3\alpha n_e s_x^{\rm{eq}}/4\hbar |G| a$. Using data of Ref.~\cite{7} ($n_e=0.7 \cdot 10^{11} \rm{cm}^{-2},$ $a=80  \rm{{\AA}},$ $\alpha=1,5 \cdot 10^{-23} \rm{eV cm}^3, $ $|G| \approx 0.2$  $s_x^{\rm{eq}} \approx 0.2$) we find $\gamma_{\rm{max}} \approx 2\cdot 10^{9} \rm{s}^{-1}.$

    Another interesting phenomenon arising due to simultaneous presence of two types of interaction
    is a  change in a sign  of the group velocity of
          the ion mode.
         It can by understood by analyzing
         the spectrum in the limit $k\to 0,$  when
           $\omega_{1,2}(k) \approx \omega_{1,2}(0) +  v_F^2k^2/2\beta.$
  Here $\beta = \Omega_e^0 [\omega_{1,2}(0) -\Omega_e^0][(\omega_{1,2}(0)-\Omega_e)^2+\delta^2]/\omega_{1,2}(0)\delta^2,$ so
 that the dispersions of the modes  are   controlled by signs of $\omega_{1}(0) -\Omega_e^0$ and $\omega_{2}(0) -\Omega_e^0$, respectively.
       As seen from Fig~1, there is a critical filed  $B =  B_0,$ at which  $\omega_1(0) = \Omega_e^0$.
            For $B <  B_0,$   both spin-wave branches are below the Stoner continuum and have  negative dispersions.
     While $B$ increases, approaching $ B_0$, the ion spin-wave branch  becomes shorter ($k_m \to 0$) and disappears when $B =  B_0$.
      For $B >  B_0,$  this  branch appears  again above the Stoner continuum and has a positive dispersion as shown in Fig.~2b.
        The dispersion of one of the  modes can also change sign for  $\Omega_e >\Omega_J$, provided that $\delta/\Omega_e \gg |G|.$

Above we discussed the coupled collective modes. Now we notice that Eqs.~\eqref{J} have also a  solution $s=0$ and $\bar{J}=0.$ Importantly, the latter equation apart from the  trivial solution $\mathbf J(\mathbf r,z,t)=0$  has also  non-zero solutions  obeying the constraint   $\int dz |\Psi(z)|^2 \mathbf J(\mathbf r,z,t)=0. $   Such solutions  were called "decoupled"   modes \cite{7}.  To find a number of such modes one should go beyond continuous approximation  and  replace integration over $dz$  in all above equations  with the sum over $N$ atomic layers. This yields $N-1$
  decoupled modes, corresponding to  independent solutions of the equation $\sum_{m} J_m  |\Psi(z_m)|^2 =0,$ where $m$ numerates layers \cite{7}.
All these modes are, indeed, decoupled from electron system provided that one neglects  the equilibrium electron exchange field acting on the ion spins.
In
this approximation, we find from Eq.~\eqref{dJdt} that the  modes  have  no dispersion and their frequencies coincide and are equal to $\Omega_J.$
In fact, weak interaction with the electrons gives rise to a small splitting of the  ion Zeeman energies, which become dependent on the  atomic layer number
$m$: $\hbar\Omega_{Jm} = \hbar\Omega_J + \alpha n_e  s_x^{\text{eq}} |\Psi(z_m)|^2$.   Taking into account this splitting, one finds  that
in a symmetric quantum well, which we  consider here, the decoupled modes with antisymmetric
distribution of ion spins $J_m \propto \delta_{m,m_0}-\delta_{m,-m_0}$ still obey the condition $\bar{J}=0,$ thus having no dispersion. For $m_0-$th mode, the ion spin precession frequency  is equal to $\Omega_{Jm_0}.$
As for the modes with a symmetric distribution, they   become  weakly coupled  to the  electron collective mode. One can show, however,
that the corresponding dispersion is very weak  provided that $n_e/n_J \ll 1.$
 Symmetric modes  also become weakly coupled to the single electron excitations and, consequently,  slowly decay in the region of
 the Stoner continuum with the characteristic   rate   $\delta^2v_F\sqrt{k^2-k_m^2}/\Omega_e\Omega_J N.$

To conclude, we have developed a theory of the spin waves in  the 2D diluted magnetic semiconductors. We have described analytically   two collective modes corresponding to  the  coupled    excitations  of the  electron and ion  spins,     and  a large number of decoupled  excitations of the ion spins. Our main finding is the tunable  anticrossing of   the collective  modes.  We  have also predicted a field-induced  change in a sign   of the group velocity of the ion mode   and  have calculated the  decay of the waves in the Stoner continuum.

We thank M.Vladimirova and D.Scalbert for valuable discussions. The work was  supported by RFBR,  by programmes of the RAS and by grant of RosNauka (project number 02.740.11.5072).


\vspace{-5mm}
\begin{thebibliography}{8}
\vspace{-5mm}
\bibitem{1}
{\it Semiconductor Spintronics and Quantum Computation}, eds. D.D.~Awschalom, D.~Loss, and N.~Samarth, in the series {\it Nanoscience and technology}, eds. K.~von~Klitzing, H.~Sakaki, and R.~Wiesendanger (Springer, Berlin, 2002).

\bibitem{2} J.~Cibert and  D.~Scalbert, {\it Diluted Magnetic Semiconductors: Basic Physics and  Optical Properties},  in {\it Spin Physics in Semiconductors}, chap. 13, ed. by M.I.~Dyakonov (Berlin, Springer, 2008).

\bibitem{3} J.A.~Gaj, R.~Planel, and G.~Fishman, Solid\ State\  Communication\ {\bf 29}, 435 (1979).

\bibitem{4} A.~Lema$\hat{\rm i}$tre, C.~Testelin,  C.~Rigaux,
T.~Wojtowicz, and G.~Karczewski, Phys.\ Rev.\ B\ {\bf 62}, 5059, (2000).

\bibitem{5}
B.~Jusserand, F.~Perez, D.R.~Richards, G.~Karczewski, T.~Wojtowicz, C.~Testelin, D.~Wolverson, and J.J.~Davies,
{ Phys.\ Rev.\ Lett.\ } {\bf 91}, 086802 (2003).

\bibitem{6} F.J.~Teran, M.~Potemski, D.K.~Maude, D.~Plantier, A.K.~Hassan,  A.~Sachrajda,
Z.~Wilamowski, J.~Jaroszynski, T.~Wojtowicz, and G.~Karczewski,  Phys.\ Rev.\ Lett.\ {\bf 91}, 077201 (2003).



\bibitem{7}
M.~Vladimirova, S.~Cronenberger, P.~Barate,  D.~Scalbert,
F.J.~Teran,
A.P.~Dmitriev,  Phys.\ Rev.\ B\ {\bf 78}, 081305(R) (2008).



\bibitem{10} F.~Perez,  { Phys. Rev. B} {\bf 79},  045306 (2009).

\bibitem{8} J.~Konig and A.H.~MacDonald { Phys.\ Rev.\ Lett.\ } {\bf 91}, 077202 (2003).

\bibitem{mm} The zero-field case was also discussed, see
D.~Frustaglia, J.~Konig, and A.H.~MacDonald, Phys.\ Rev.\ B\ {\bf 70},
045205 (2004).




\bibitem{11} P.M.~Platzmann and E.~Wolf, {\it Waves and Interaction in Solid State Plasmas}  (Academic, New York 1973).



\bibitem{12}
    A.I.~Akhiezer, V.G.~Baryakhtar and S.V.~Peletminskii, {\it Spin Waves} (North-Holland, Amsterdam, 1968).
\bibitem{13} For low temperature,
one can search the  electron spin density in the following form $\mathbf s(\mathbf p,\mathbf r,t)=n_e\nu^{-1}\delta(\epsilon - E_F)\mathbf s(\mathbf r,\varphi,t).$

\end{thebibliography}
\end{document}